\documentclass{ws-procs975x65}
\usepackage{graphics}
\usepackage{color}
\usepackage{amsfonts}
\usepackage{latexsym}
\usepackage{enumerate}
\providecommand{\U}[1]{\protect\rule{.1in}{.1in}}

\DeclareMathOperator{\arcsinh}{arcsinh}

\newcommand{\be}{\begin{equation}}
\newcommand{\ee}{\end{equation}}
\newcommand{\bea}{\begin{eqnarray}}
\newcommand{\ea}{\end{eqnarray}}
\newcommand{\bean}{\begin{eqnarray*}}
\newcommand{\eean}{\end{eqnarray*}}
\def\bal#1\eal{\begin{align}#1\end{align}}
\begin{document}
\title{Quantization of the Szekeres spacetime through generalized symmetries}
\author{Andronikos Paliathanasis}
\address{Instituto de Ciencias F\'{\i}sicas y Matem\'{a}ticas, Universidad Austral de
Chile, Valdivia, Chile}
\address{Institute of Systems Science, Durban University of Technology, POB 1334
Durban 4000, South Africa.}
\author{Adamantia Zampeli \footnote{azampeli@phys.uoa.gr}}
\address{Institute of Theoretical Physics, Faculty of Mathematics and Physics,
Charles University, \\
V Hole\v{s}ovi\v{c}k\'ach 2, 18000 Prague 8, Czech
Republic}
\author{Theodosios Christodoulakis}
\address{Nuclear and Particle Physics section, Physics Department, University of
Athens, 15771 Athens, Greece}
\author{M.T. Mustafa}
\address{Department of Mathematics, Statistics and Physics, College of Arts and
Sciences, Qatar University, Doha 2713, Qatar}
\begin{abstract}
We present the effect of the quantum corrections on the Szekeres spacetime, a system important for the study of the inhomogeneities of the pre-inflationary era of the universe. The study is performed in the context of canonical quantisation in the presence of symmetries. We construct an effective classical Lagrangian and impose the quantum version of its classical integrals of motion on the wave function. The interpretational scheme of the quantum solution is that of Bohmian mechanics, in which one can avoid the unitarity problem of quantum cosmology. We discuss our results in this context.
\end{abstract}

\keywords{Szekeres system; Silent universe; Quantisation; Semiclassical approach}
\bodymatter
\section{Introduction}
We focus on the quantisation of the Szekeres spacetime metric with the aim to study the effect of possible quantum corrections in the dynamics. This metric has the form \cite{Szekeres:1974ct}
\begin{equation}
ds^{2}=-dt^{2}+e^{2\alpha}dr^{2}+e^{2\beta}\left( dy^{2}+dz^{2}\right)
\label{ss.04aa}
\end{equation}
where $\alpha \equiv \alpha\left( t,r,y,z\right) $ and $\beta
\equiv\beta\left( t,r,y,z\right)$ and represents an irrotational perfect fluid with vanishing pressure and magnetic Weyl tensor, $p=\omega_{ab} =H_{ab}=0$.
The interest in the silent universe lies on the fact that it can be seen as inhomogeneous solutions of Einstein equations with no symmetries which generalise Kantowski-Sachs, FRW and Tolman-Bondi spacetimes. Thus it is proper for the description of perturbations on these spacetimes \cite{Ishak:2011hz,Bolejko:2010eb,Vrba:2014dwa,Bruni:1994nf}.


We start by writing the field equations on the covariant variables $(\rho, \theta, \sigma, E)$, where $\rho=T^{\mu\nu }u_{\mu}u_{\nu}$, with $T^{\mu\nu}$ being the energy-momentum tensor of the matter, $\theta=\left( \nabla_{\nu}u_{\mu}\right) h^{\mu\nu}$ is
the expansion rate of the observer, while $\sigma~$and $E$ are the shear
and electric component of the Weyl tensor, $E_{\nu}^{\mu}=$ $Ee_{\nu}^{\mu},~\sigma_{\nu}^{\mu }=\sigma e_{\nu}^{\mu},$ in which the set
of $\left\{ u^{\mu} ,e_{\nu}^{\mu}\right\} $ defines an orthogonal tetrad. The field equations then become
\begin{subequations}\label{szeksys}
\begin{align}
&\dot{\rho}+\theta\rho =0,~  \label{ss.01} \\
&\dot{\theta}+\frac{\theta^{2}}{3}+6\sigma^{2}+\frac{1}{2}\rho =0,
\label{ss.02} \\
&\dot{\sigma}-\sigma^{2}+\frac{2}{3}\theta\sigma+E =0,  \label{ss.03} \\
&\dot{E}+3E\sigma+\theta E+\frac{1}{2}\rho\sigma =0,  \label{ss.04}
\end{align}
together with the algebraic equation which is the Hamiltonian constraint
\begin{equation}
\frac{\theta^{2}}{3}-3\sigma^{2}+\frac{^{\left( 3\right) }R}{2}=\rho
\end{equation}
\end{subequations}
where $\dot{}$ denotes the directional derivative along $u^{\mu}$, the energy density is $\rho=T^{\mu\nu }u_{\mu}u_{\nu}$, with $T^{\mu\nu}$ being the energy-momentum tensor of the matter, the parameter $\theta=\left( \nabla_{\nu}u_{\mu}\right) h^{\mu\nu}$ is
the expansion rate of the observer, while $\sigma~$and $E$ are the shear
and electric component of the Weyl tensor, $E_{\nu}^{\mu}=$ $%
Ee_{\nu}^{\mu},~\sigma_{\nu}^{\mu }=\sigma e_{\nu}^{\mu},$ in which the set
of $\left\{ u^{\mu} ,e_{\nu}^{\mu}\right\} $ defines an orthogonal tetrad. We note that in addition to these equations, the spatial constraints ensure the integrability of the system.

In the following sections we present the quantization of this system in terms of canonical quantization in the presence of symmetries \cite{Christodoulakis:2012eg}. The starting point is the effective Lagrangian obtained in \cite{Paliathanasis:2017wli} and we adopt the Bohmian approach for our analysis \cite{Bohm:1951xw,Bohm:1951xx} following e.g. \cite{Zampeli:2015ojr}, since its causal character suits the context of quantum cosmology, where the notion of an external observer cannot be justified.
\section{Classical and Quantum Dynamics}
In \cite{Paliathanasis:2017wli} the Szekeres system \eqref{szeksys} was written in an
equivalent form of a two second-order differential equations system which are equations of motion of a Lagrangian of the form
\be 
L= \frac{1}{2} G_{\alpha\beta} (q(t))\dot{q}^\alpha(t) \dot{q}^\beta (t)- V(q(t)), \quad \alpha,\beta= 0,\ldots n-1
\ee
where $q(t)$ denote the degrees of freedom of the system and $G_{\alpha \beta}$ the metric on the configuration space of variables. Adopting proper coordinates for our case, $\rho = \frac{6}{\left( 1-v  \right) u^2}, \ E= \frac{v}{u^3 \left( v-1 \right)}$, the system of the two second order equations becomes
\begin{subequations}\label{szeksysred}
\bal
&\ddot{v} - \frac{2v}{u^3}=0, \\
&\ddot{u} + \frac{1}{u^2} = 0 \label{e-l2}
\eal
\end{subequations}
derivable from $L= \dot{u} \dot{v} -\frac{\dot{v}}{u^2}$. The system \eqref{szeksysred} admits two integrals of motion, quadratic in the velocities; the first is the Hamiltonian function since the system is autonomous, while the second one is the quadratic function $I_0$ which can be constructed by the application of Noether's theorem for contact symmetries \cite{Paliathanasis:2017wli}, which in the phase space become
\begin{subequations}
\begin{align}
& p_{u}p_{v}+\frac{v}{u^{2}}=h,  \label{sss.30} \\
& p_{v}^{2}-2u^{-1}=I_{0}  \label{cons0}
\end{align}%
\end{subequations}
When turned to quantum operators and imposed on the wave function, according to the rules $\hat{p}_\alpha = - i \frac{\partial}{\partial q^\alpha} , \quad \{.,.\} \rightarrow -\frac{i}{\hbar} [.,.]$, with operator-ordering respecting the general covariance and hermiticity ensured under the inner product $\int d^n q \mu \psi^*_1 \psi_2$, they  lead to two eigenvalue equations
\begin{subequations}
\label{sss.300}
\begin{align}
&\left( -\partial _{uv}+\frac{v}{u^{2}}\right) \Psi =h\Psi ,  \label{sss.31}\\
&\left( \partial _{vv}+\frac{2}{u}\right) \Psi =-I_{0}\Psi ,  \label{sss.32}
\end{align}
\end{subequations}
In the first one we can recognise the time-independent Schro\"odinger equation and their solution is
\begin{equation}
\Psi \left( I_{0},u,v\right) =\frac{\sqrt{u}}{\sqrt{2+I_{0}u}}\left( \Psi
_{1}\cos f\left( u,v\right) +\Psi _{2}\sin f\left( u,v\right) \right)
\label{generalsol}
\end{equation}%
where~%
\begin{align}
& f\left( u,v\right) =\frac{(hu+I_{0}v)\sqrt{2I_{0}+I_{0}^{2}u}-2h\sqrt{u}%
\arcsinh\sqrt{\frac{I_{0}{u}}{2}}}{I_{0}^{3/2}\sqrt{u}},\quad \text{\ for }%
I_{0}\neq 0,  \label{sss.33} \\
&  \notag \\
& f\left( u,v\right) =\frac{\sqrt{2}\left( hu^{2}+3v\right) }{3\sqrt{u}}%
,\quad \text{for }I_{0}=0.  \label{sss.344}
\end{align}%
and $\Psi _{1}, \ \Psi _{2}$ denote constants of integration. Due to the linearity of \eqref{sss.31}, the general solution is $\Psi _{Sol}\left(
u,v\right) =\sum_{I_{0}}\Psi \left( I_{0},u,v\right) $.
\section{Semiclassical analysis and probability}
In the context of the Bohmian approach, the departure from the classical theory is determined by an additional term in the classical Hamilton-Jacobi equation, known as quantum potential $Q_{V}=-\frac{\Box \Omega }{2\Omega }$, where $\Omega $ denotes the amplitude of the wave function in polar form, $\Psi (u,v)=\Omega (u,v)e^{iS(u,v)}$. When the quantum potential is zero, the identification
\begin{equation}  \label{semiclas}
\frac{\partial S}{\partial q_i}=p_{i}=\frac{\partial }{\partial\dot{q}_{i}}
\end{equation}
is possible. If this classical definition for the momenta is retained even when $Q\neq0$, the semiclassical solutions will differ from the classical ones. 

\begin{figure}[ptb]
\includegraphics[height=3.5cm]{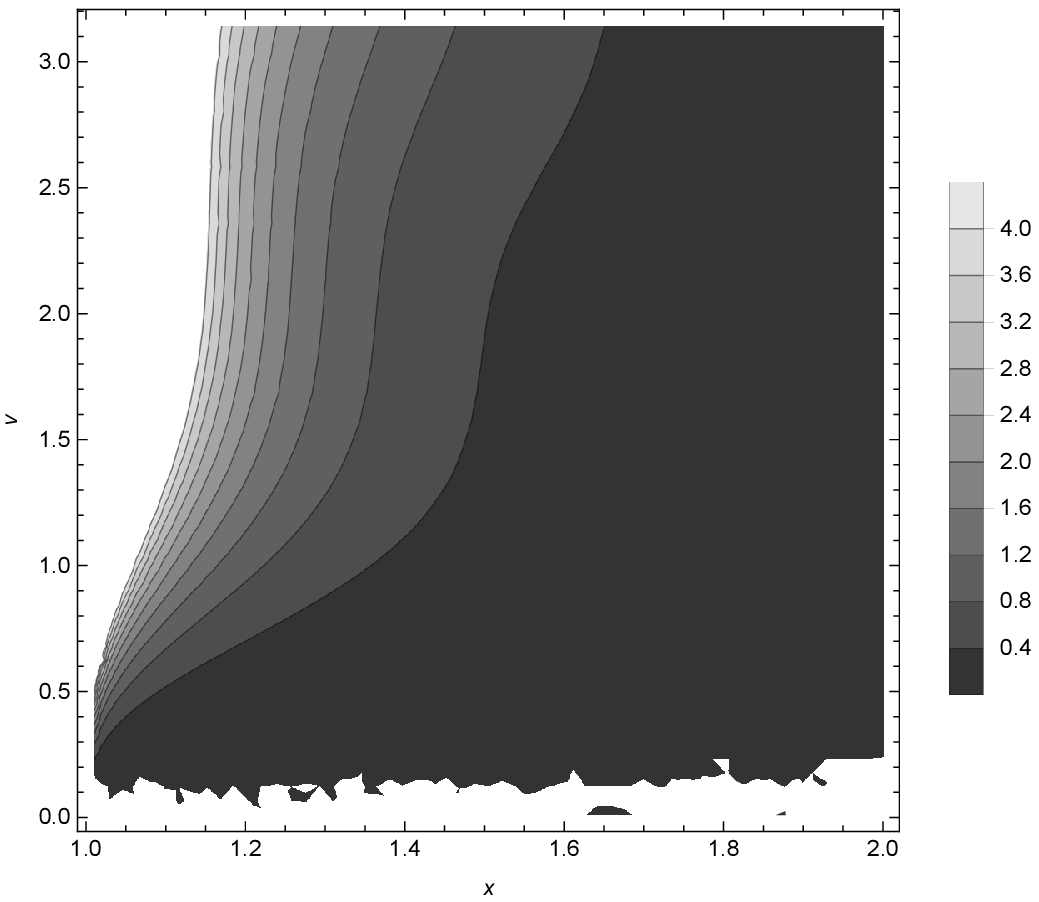}\centering
\includegraphics[height=3.5cm]{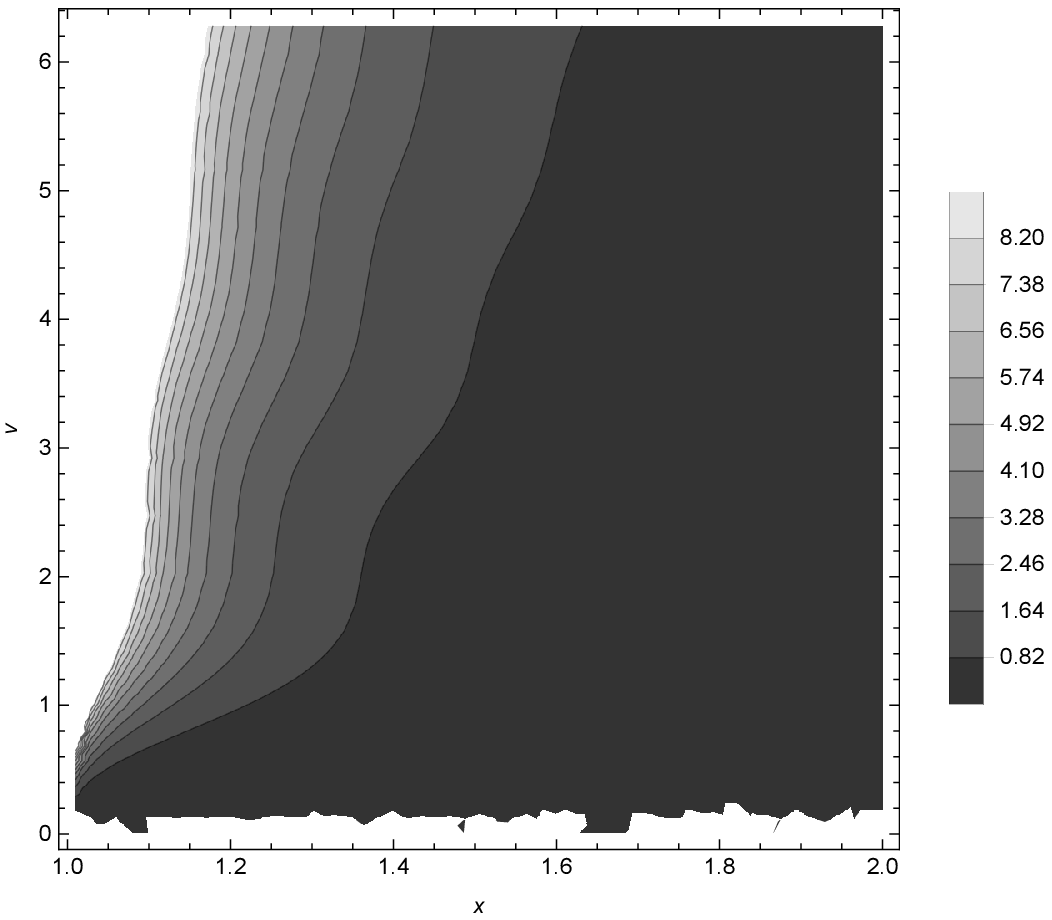}\centering
\caption{Contour plot for the probability function in the
space of variables ${x,v}$. We observe that as $x\rightarrow0$ and $v$ is
small, that is, $I_{0}\rightarrow0$, the function $P\left( x,v\right) $
reaches to a minimum extreme.}\label{plot333}
\end{figure}

Under the assumption that the quantum corrections in the general solution \eqref{generalsol} follow from the ``frequency $I_{0}$" with the highest
peak in the wave function, which is in agreement with the so-called Hartle
criterion \cite{Hartle1987}, the quantum potential
vanishes. This provides no quantum corrections and the semiclassical equations \eqref{semiclas} give the classical solution, i.e. the Szekeres universe remains ``silent", even at the quantum level. In the particular case $h=0$ and $\Psi_1 \rightarrow 0$, the wave function is well behaved at $u\rightarrow 0$ and $u\rightarrow \infty $. We can thus define a probability which, after a change of coordinates to $u\rightarrow \frac{2}{x^{2}-I_{0}}$ becomes
\begin{equation}
P=\int_{\sqrt{I_{0}}+\epsilon }^{\lambda }dx\int_{0}^{2k\pi }dv\ \frac{%
4c_{3}^{2}\sin (xv)}{x(x^{2}-I_{0})^{2}}~,~k\in \mathbb{N}.  \label{in1}
\end{equation}
where the cut-off constant $\lambda$ is introduced to exclude the case $E=0,\rho =0$. The qualitative behaviour of the probability function is given in the contour plot in Fig. \ref{plot333}. The plots show that for $I_{0}\rightarrow 0$ the probability function reaches its minimum.
\section{Conclusions}
Our quantum analysis of the Szekeres system was based on the canonical quantization in the presence of symmetries and the results were interpreted by adopting the Bohmian mechanics approach. The starting point was an effective classical point-like Lagrangian which can reproduce the two dimensional system of second-order differential equations resulted from the initial field equations. This Lagrangian is
autonomous, thus there exists a conservation law of ``energy" corresponding to the Hamiltonian function. As for the extra contact symmetry, it leads to a quadratic in the momenta conserved quantity attributed to a Killing tensor of the second-rank. The two conserved quantities give two eigenequations at the quantum level, the Hamiltonian function being the Schr\"{o}dinger equation.

The assumption that the wave function is peaked around its classical value leads to the lack of quantum corrections and the recovery of the classical solutions, thus leading to the conclusion that the Szekeres universe remains silent at the
quantum level. Finally, for the particular case $h=0$ it was shown that the probability function and relate one (unstable) exact solution with the
existence of a minimum of this probability. 

The classical exact solution which corresponds at the values $h=I_{0}=0$ of the integration constants is $u_{A}\left( t\right) =\frac{6^{\frac{2}{3}}}{2}t^{\frac {2}{3%
}},v_{A}\left( t\right) =v_{0}t^{-\frac{1}{3}}$ \cite{Paliathanasis:2017wli} and corresponds to an unstable critical point for the dynamical system %
\eqref{szeksys}. It is interesting the fact that these conditions also correspond to the extremum of the probability function, something which might be related to the existence and stability of the exact solution. This result is also in accordance with the
analysis of the probability extrema in \cite{Dimakis:2016mpg} where it was shown that
the extrema of the probability lie on the classical values.
\bibliographystyle{ws-procs975x65}
\bibliography{szekeres}

\end{document}